\documentclass[preprint,12pt,sort&compress]{elsarticle}
 
\usepackage{graphicx}
\usepackage{amsmath}
\usepackage{amssymb}
\usepackage[section]{placeins}

\usepackage{xcolor}
\colorlet{linkequation}{red}
\usepackage[colorlinks]{hyperref}

\begin{document}
\begin{frontmatter}

\title{The {\bf S3} Symmetric Model with a Dark Scalar}

\author[conaIF]{C. Espinoza\corref{cor1}}
\ead{m.catalina@fisica.unam.mx}
\cortext[cor1]{Corresponding author}

\author[IF]{E. A. Garc\'{e}s}
\ead{egarces@fis.cinvestav.mx}

\author[IF]{M. Mondrag\'{o}n}
\ead{myriam@fisica.unam.mx}

\author[CNRS]{H. Reyes-Gonz\'{a}lez}
\ead{Humberto.Reyes-Gonzalez@lpsc.in2p3.fr}

\address[conaIF]{C\'atedras CONACYT - Instituto de F\'{i}sica, Universidad Nacional Aut\'{o}noma de M\'{e}xico, Apdo. Postal 20-364, Cd. M\'{e}xico 01000, M\'{e}xico.}


\address[IF]{Instituto de F\'{i}sica, Universidad Nacional Aut\'{o}noma de M\'{e}xico, Apdo. Postal 20-364, Cd. M\'{e}xico 01000, M\'{e}xico.}

\address[CNRS]{Laboratoire de Physique Subatomique et de Cosmologie, Universit\'e Grenoble-Alpes, CNRS/IN2P3, 53 Avenue des Martyrs, F-38026 Grenoble, France.}

\begin{abstract}
We study the $S3$ symmetric extension of the Standard Model in which all the irreducible representations of the permutation
group are occupied by $SU(2)$ scalar doublets, one of which is taken as inert and can lead to dark matter candidates. We perform a scan over parameter space probing points
against physical constraints ranging from unitarity tests to experimental Higgs searches limits. We find that the latter constraints severely restrict 
the parameter space of the model. For acceptable points 
we compute the value of the relic density of the dark scalar candidates  
and find that it has a region for low dark matter masses which complies with the Higgs searches bounds and lies within the experimental Planck limit.  For masses $\gtrsim 80$ GeV the value of the relic density is below the Planck bound, and it reaches values close to it for very heavy masses $\sim 5$ TeV.  In this heavy mass region, this opens the interesting possibility of extending the dark sector of the model with additional particles.

\end{abstract}

\begin{keyword}



Dark matter \sep multi-Higgs model \sep $S3$ symmetric model \sep \href{https://arxiv.org/abs/1804.01879}{arXiv:1804.01879}

\end{keyword}

\end{frontmatter}


\section{Introduction}
\label{intro}

One of the main  challenges in particle physics at present is to find the
nature of dark matter (DM). It is generally accepted that it should be
neutral, cold and weakly interacting (although other possibilities
exist), and there are proposals for scalar, fermion and vector
particles that satisfy the criteria~\cite{Taoso:2007qk}.  Among the scalar candidates a
very interesting proposal is to introduce inert Higgs scalars,
i.e. that do not couple to matter, which is usually achieved by
introducing an extra $Z_2$ discrete symmetry,  and that 
do not acquire a vacuum expectation value (vev),
thus guaranteeing the stability of the DM candidate;
for a single inert scalar
this model is referred to in the literature as Inert Doublet Model
(IDM or i2HDM)~\cite{Belyaev:2016lok,Ginzburg:2010wa,Barbieri:2006dq,LopezHonorez:2006gr,LopezHonorez:2010tb,Hambye:2007vf,Hambye:2009pw,Honorez:2010re,McDonald:1993ex,Hirsch:2010ru}.

Among the numerous proposals to extend the scalar sector of the
standard model, the 3-Higgs Doublet Model (3HDM) with an $S_3$-family
symmetry (S3-3H) presents interesting phenomenology, such as the
prediction of a non zero reactor neutrino mixing angle $\theta_{13}$
and of a CKM matrix in accordance with the experimental results
\cite{Canales:2013cga}.  The 3HDM with S3 symmetry has been extensively studied in
different contexts (see for instance, \cite{Mondragon:1998ir, Kubo:2003iw,Mondragon:2007af,
Canales:2012dr,Canales:2013cga,Das:2014fea, 
Barradas-Guevara:2014yoa,Emmanuel-Costa:2016vej}, and
references therein).  The
aim of this project is to extend this model to a 4HDM in order to have
dark matter candidates, without spoiling the good features of the
S3-3H model in the quark and lepton sectors. To to do this we occupied all irreducible
representations of the $S_3$ symmetry: one symmetric singlet, one
antisymmetric singlet and one doublet. The S3-3H model is constituted by the
singlet symmetric and doublet representations, with all these
Higgs scalars acquiring vacuum expectation values. The fourth
Higgs doublet is assigned to the antisymmetric singlet representation 
and assumed to be inert, it does not acquire a vev and we impose a $Z_2$ 
symmetry to ensure the stability of the potential dark matter
candidates.

In this letter we present an analysis of the parameter space of the model focusing in the
determination of the relic density of the dark scalar for physically acceptable points by requiring 
them to satisfy numerous physical constraints.
In the study presented here we limit the calculations to include only tree level quantities, for example the values of
the quartic couplings are approximated by on shell values.
While higher order corrections can be of sizeable importance for non-supersymmetric models 
(see e.g. \cite{Krauss:2017xpj,Braathen:2017jvs}), an analysis including
full loop corrections is outside the scope of this work, and we leave it for future research.

\section{The model}
\label{model}

In the $S3$ symmetric model the scalar sector of the SM is extended with additional
$SU(2)$ scalar doublets $\Phi_k$ with definite transformation properties with respect to the permutation symmetry.
Whilst the matter sector content of the model remains the same as the SM, the Yukawa lagrangian is required to be invariant
also with respect to  $S3$. This, together with the mixing of the scalars after electroweak symmetry breaking,
leads to Yukawa terms that can be very different from the SM, for example the proportionality of the
fermion masses to single Yukawa couplings is in general lost in the extended model. Since our primary focus in this
letter will be centred in the properties of the dark scalar, in particular a probe for the values of the relic density
of this particle in the model parameter space, we'll make simplifying assumptions over the Yukawa lagrangian's explicit form
which we shall argue not to have a strong impact in the conclusions driven from our results.

\subsection{The scalar sector}

We accommodate four $SU(2)$ doublets into the irreducible representations of the permutation group $S3$, 
denoting the symmetric and antisymmetric scalars by $\Phi_s$ and $\Phi_a$ respectively, while the remaining
two doublet scalars $\Phi_1$ and $\Phi_2$ are arranged in a column vector conforming the $S3$ doublet.
The (invariant and renormalizable) scalar potential is a mixture of the potentials known from the studies of the three Higgs model with
the permutation symmetry (see for instance \cite{Kubo:2004ps,Beltran:2009zz,Das:2014fea,Barradas-Guevara:2014yoa,Emmanuel-Costa:2016vej}), and can be written as:

\begin{equation}
\label{pot}
V = V_2 + V_4 + V_{4s} + V_{4a} + V_{4sa},
\end{equation}
where $V_2$ comprises the quadratic terms:

\begin{equation}
\label{pot2}
V_2 = \mu_{0}^{2}\Phi_{s}^{\dagger}\Phi_{s}+\mu_{1}^{2}(\Phi_{1}^{\dagger}\Phi_{1}+\Phi_{2}^{\dagger}\Phi_{2})+\mu_{2}^{2}\Phi_{a}^{\dagger}\Phi_{a},
\end{equation}
$V_4$ contains quartic terms involving $\Phi_1$ and $\Phi_2$ only:

\begin{eqnarray}
\label{pot4}
V_4 & = &  \lambda_{1}(\Phi_{1}^{\dagger}\Phi_{1}+\Phi_{2}^{\dagger}\Phi_{2})^{2}+\lambda_{2}(\Phi_{1}^{\dagger}\Phi_{2}-\Phi_{2}^{\dagger}\Phi_{1})^{2} \nonumber
\\ &  & +\lambda_{3}[(\Phi_{1}^{\dagger}\Phi_{1}-\Phi_{2}^{\dagger}\Phi_{2})^{2}+(\Phi_{1}^{\dagger}\Phi_{2}+\Phi_{2}^{\dagger}\Phi_{1})^{2}],
\end{eqnarray}
while $V_{4s}$ and $V_{4a}$ represent the quartic terms involving $\Phi_s$ and $\Phi_a$ respectively:

\begin{eqnarray}
\label{pot4s}
V_{4s} & = & \lambda_{4}[(\Phi_{s}^{\dagger}\Phi_{1})(\Phi_{1}^{\dagger}\Phi_{2}+\Phi_{2}^{\dagger}\Phi_{1})+(\Phi_{s}^{\dagger}\Phi_{2})(\Phi_{1}^{\dagger}\Phi_{1}-\Phi_{2}^{\dagger}\Phi_{2})+\mathrm{h.c.}]  \nonumber
\\ &  & +\lambda_{5}(\Phi_{s}^{\dagger}\Phi_{s})(\Phi_{1}^{\dagger}\Phi_{1}+\Phi_{2}^{\dagger}\Phi_{2}) \nonumber
\\ &  & +\lambda_{6}[(\Phi_{s}^{\dagger}\Phi_{1})(\Phi_{1}^{\dagger}\Phi_{s})+(\Phi_{s}^{\dagger}\Phi_{2})(\Phi_{2}^{\dagger}\Phi_{s})] \nonumber
\\ &  & +\lambda_{7}[(\Phi_{s}^{\dagger}\Phi_{1})(\Phi_{s}^{\dagger}\Phi_{1})+(\Phi^{\dagger}_{s}\Phi_{2})(\Phi^{\dagger}_{s}\Phi_{2})+\mathrm{h.c.}] \nonumber
\\ &  & +\lambda_{8}(\Phi^{\dagger}_{s}\Phi_{s})^{2} 
\end{eqnarray}
The expression for $V_{4a}$ is very similar to eq. (\ref{pot4s}) with $\Phi_a$ replacing $\Phi_s$ and quartic couplings 
$\lambda_9, \ldots , \lambda_{13}$, except that the $\lambda_9$ term analogous to the $\lambda_4$ term has $\Phi_1$ and $\Phi_2$
interchanged. Finally the mixed $\Phi_s, \Phi_a$ term is given by:

\begin{equation}
V_{4sa} = \lambda_{14} (\Phi_{s}^{\dagger} \Phi_{a}   \Phi_{a}^{\dagger} \Phi_{s}) + 
          \lambda_{15} (\Phi_{1}^{\dagger} \Phi_{s} \Phi_{2}^{\dagger} \Phi_{a} + \mathrm{h.c.}) 
\end{equation}

In the following we will assume no additional sources of CP violation
other than those present in the SM and hence we shall take the quartic
couplings $\lambda_i$, $i=1 \ldots 14$ to be real and also restrict
their absolute values with the usual perturbativity condition
$|\lambda_i|<4\pi$. In order to force the scalar $\Phi_a$ to be inert
we introduce an additional discrete $Z_2$ symmetry with respect to
which all fields are even except those with subindex $a$, taken as
$Z_2$-odd.  This gets rid of the $\lambda_9$ and  $\lambda_{15}$ term in the potential
leaving only terms with even powers of $\Phi_a$. Incidentally this
also leads to the appearance of an additional symmetry of the
potential under the interchange $\Phi_1 \rightarrow -\Phi_1$, this
fact explains the absence of vertices with odd number of fields with
subindex $1$ in the Feynman rules.

After electroweak symmetry breaking all the scalar doublets acquire a
vacuum expectation value (vev) denoted by $v_s$, $v_1$, $v_2$ and
$v_a$ respectively. However, in order to avoid the explicit breaking
of the $Z_2$ symmetry we fix $v_a=0$ and henceforth from the
minimization conditions for the scalar potential (tadpole equations)
the fourth equation $\partial V / \partial v_a = 0$ is automatically
satisfied
\footnote{Many of the expressions presented here can be obtained
  using \texttt{SARAH}~\cite{Staub:2013tta,Staub:2009bi,Staub:2010jh,Staub:2012pb}; the model files
	can be downloaded from the \texttt{SARAH} model database~\cite{SARAH-DB}.}. 
The three minimization equations 
($\partial V / \partial v_i = 0$, $i$ = s, 1, 2) reduce to those of
the three Higgs doublet model with $S3$ symmetry (e.g. \cite{Barradas-Guevara:2014yoa,Das:2014fea}),

\begin{align}
\begin{aligned}
\label{murel}
\mu^{2}_{0} &= -\frac{1}{2}(\lambda_5 + \lambda_6 + 2\lambda_7)(v_1^2 + v_2^2) -  \lambda_8 v_s^2 +\frac{\lambda_4(v_2^2 - 3v_1^2)v_2}{2v_s}
\\
\mu^{2}_{1} &= -\frac{1}{2}(\lambda_5 + \lambda_6 + 2\lambda_7)v_s^2 - (\lambda_1 + \lambda_3)(v_1^2 + v_2^2) - 
  3\lambda_4 v_2 v_s
  \\
  \mu^{2}_{1} &= -\frac{1}{2}(\lambda_5 + \lambda_6 + 2\lambda_7)v_s^2 - (\lambda_1 + \lambda_3)(v_1^2 + v_2^2) +
  \frac{3\lambda_4 v_{s}(v_2^2-v_1^2)}{2v_2}
  \end{aligned}
\end{align}
whose {more general CP preserving solution is}
$v_1 = \sqrt{3} v_2$, 
 with the usual SM vev
given by $v = \sqrt{v_s^2 + 4 v_2^2} = \mathrm{246 \, GeV}$.
{It is convenient to parametrize in spherical coordinates, 
\begin{align}\label{spherparam}
\begin{aligned}
\begin{split}
v_s={}v\cos\theta,~v_1={}v\sin\theta\cos\phi, ~v_2={}v\sin\theta\sin\phi,
\end{split}
\end{aligned}
\end{align} 
where $\theta \ \in \ (0,\pi)$ and $\phi \ \in \ (0,2\pi)$. 
With this parametrization we get $\tan^{2}\phi=\frac{1}{3}$ and $v_s=v\cos\theta$, $v_2=\frac{1}{2}v\sin\theta$.} 
{We choose $\tan\theta = 2 v_2/v_s$ as one of the independent parameters in the numerical calculations.}

Parametrizing the Higgs doublets as

\begin{equation}\label{param}
\Phi_s = \left( \begin{array}{c}
         h^{\prime +}_s \\
         (v_s + h^{\prime n}_s + i h^{\prime p}_s)/\sqrt{2} \end{array} \right)
\end{equation}
and similarly for $\Phi_1$, $\Phi_2$ and $\Phi_a$ (here the indices $n$ and $p$ refer to {\it neutral scalar} and {\it pseudoscalar} respectively, 
and we use primes to distinguish from the mass eigenstates which we will denote with the same letters except when explicitly stated otherwise) it is straightforward to obtain the mass and mixing matrices and
mass eigenstates for the scalar fields. 
{The $Z_2$ odd fields do not mix (e.g. $h^{\prime n}_a=h^{n}_a$) and therefore the mass and mixing matrices have block diagonal form. In the case of the neutral scalar fields the submatrix mixing the $Z_2$ even fields $h^{\prime n}_s$, $h^{\prime n}_1$ and $h^{\prime n}_2$ into the mass eigenstates $H$, $H_3$ and $h$ takes the form:

\begin{equation}\label{rotM}
Z =
\left(
\begin{array}{ccc}
 \cos(\alpha) & 0 & \sin(\alpha) \\
 0 & 1 & 0 \\
 -\sin(\alpha) & 0 & \cos(\alpha) \\
\end{array}
\right)
\left(
\begin{array}{ccc}
 1 & 0 & 0 \\
 0 & \frac{1}{2} & -\frac{\sqrt{3}}{2} \\
 0 & \frac{\sqrt{3}}{2} & \frac{1}{2} \\
\end{array}
\right)
\end{equation}
so that the neutral scalar mass matrix $m_{h^n}^2$ is diagonalized by

\begin{equation}\label{zeta}
\mathrm{diag}(m^2_H, m^2_{H_3}, m^2_h) = Z m_{h^n}^2 Z^\mathrm{T}
\end{equation}
here we have resorted to a notation for the neutral scalar mass eigenstates that facilitates
comparison with analogue expressions found in the literature of the Two Higgs Doublet Model (THDM)
(see e.g. \cite{Branco:2011iw}), thus $H$ and $h$ are the physical fields encountered also in the THDM,
with the latter being the lightest of the two. On the other hand $H_3$ is an additional field not present
in THDMs and cannot be related in any form with the}
{SM Higgs, since it does not have couplings to the vector bosons~\cite{Das:2014fea}.
}
{From the first rotation in (\ref{zeta}) the submatrix ${\cal M}_{ij}$ with $i$, $j=1,2$, analogue to the mixing matrix of the THDM is found to be:

\begin{eqnarray}\label{Ms}
{\cal M}_{11} & = & \frac{1}{2} v^2 (4\lambda_8 \cos^2{\theta} - \lambda_4 \sin^2{\theta} \tan{\theta}) \nonumber \\ 
{\cal M}_{12} & = & \frac{1}{2} v^2 \sin{\theta} (2(\lambda_5 + \lambda_6 + 2\lambda_7) \cos{\theta} + 3\lambda_4 \sin{\theta}) \\ 
{\cal M}_{22} & = & \frac{1}{2} v^2 \sin{\theta} (3\lambda_4 \cos{\theta} + 4(\lambda_1 + \lambda_3) \sin{\theta})  \nonumber
\end{eqnarray}
and defines the rotation angle $\alpha$ through the expressions (e.g. see \cite{Gunion:2002zf}):

\begin{eqnarray}\label{alfa}
\sin{(2\alpha)} & = & \frac{2{\cal M}_{12}}{\sqrt{({\cal M}_{11} - {\cal M}_{22})^2 + 4({\cal M}_{12})^2}} \nonumber \\
\cos{(2\alpha)} & = & \frac{{\cal M}_{11} - {\cal M}_{22}}{\sqrt{({\cal M}_{11} - {\cal M}_{22})^2 + 4({\cal M}_{12})^2}}
\end{eqnarray}
The mass eigenvalues corresponding to $H$ and $h$ are neatly expressed in terms of the elements of ${\cal M}$:

\begin{equation}\label{mHh}
m^2_{H,h} = \frac{1}{2} \left(  {\cal M}_{11} + {\cal M}_{22} \pm \sqrt{({\cal M}_{11} - {\cal M}_{22})^2 + 4({\cal M}_{12})^2} \right)
\end{equation}
where the lower sign corresponds to $h$. The mass eigenvalue of $H_3$ is simply:

\begin{equation}\label{mH3}
m^2_{H_3} = -\frac{9}{4} \lambda_4 v^2 \sin{(2\theta)}
\end{equation}
}
{ The connection of $H$ and $h$ with the SM Higgs is done as usual through the decoupling limit defined by the relation
$\cos{(\theta - \alpha)} \approx 0$,
%
in this limit $h$ has SM-like couplings and can be identified with the SM Higgs. In the numerical calculation
we impose the mass of $h$ to be always around 125 GeV, so that for points satisfying the decoupling limit we recover
the measured mass of the scalar discovered at CERN~\cite{Aad:2012tfa,Chatrchyan:2012xdj} within experimental error bars.
}

{The dark matter candidate can be either the neutral scalar $h_a^n$ or the pseudoscalar $h_a^p$, both come from the inert doublet and have odd charges with respect to the $Z_2$ symmetry,
and their masses are found to be:

\begin{eqnarray}\label{darkmasses}
M^2_{h_a^n} & = \mu^2_2 + \frac{1}{2} v^2 (\lambda_{14} \cos^2{\theta} + \lambda_X^+ \sin^2{\theta}) & \\
M^2_{h_a^p} & = \mu^2_2 + \frac{1}{2} v^2 (\lambda_{14} \cos^2{\theta} + \lambda_X^- \sin^2{\theta}) & 
\end{eqnarray}
where the effective parameter $\lambda_X^\pm=\lambda_{10} + \lambda_{11}\pm 2\lambda_{12}$ characterizes the difference between the masses of the dark neutral particles.

The rest of the field content of the model include additional pseudo-scalar fields $A$, $h_2^p$, and additional charged fields
$H^\pm$, $h_2^\pm$ and $h_a^\pm$; $A$ and $H^\pm$ being the analogue of the fields appearing in the THDM.}
{ 
Explicit expressions for the masses of these fields are given below:

\begin{equation}\label{masses}
\begin{array}{l@{}l}
M^2_{h_a^+} &{}= \mu^2_2 + \frac{1}{2} v^2 \lambda_{10}  \sin^2{\theta}   \\
M^2_{h_2^+} &{}= -\frac{1}{4} v^2 (\lambda_{6} + 2\lambda_7 + 4\lambda_3 + (\lambda_{6} + 2\lambda_7 - 4\lambda_3)\cos{2\theta} + 5\lambda_4 \sin{2\theta})  \\ 
M^2_{H^+}   &{}=  -\frac{1}{2} v^2 (\lambda_{6} + 2\lambda_7 + \lambda_4 \tan{\theta})  \\
M^2_{A}     &{}=  -\frac{1}{2} v^2 (4\lambda_7 + \lambda_4 \tan{\theta})   \\ 
M^2_{h_2^p} &{}=  -\frac{1}{2} v^2 \sin^2{\theta} (4(\lambda_2 + \lambda_3) + 5 \lambda_4 \cot{\theta} + 4 \lambda_7 \cot^2{\theta})   \\
\end{array}
\end{equation}
}

Following the procedure outlined in \cite{Emmanuel-Costa:2016vej}, the
stability constraints for this model can be obtained (see \cite{Espinoza:2017ryu}).
The (tree level) unitarity conditions are calculated using the LQT prescription~\cite{Lee:1977eg}; we consider all possible
combination of two-particle scattering processes (including charged ones) at high energies involving 
longitudinal gauge bosons and scalars. As is well known from the Goldstone boson equivalence theorem,
these amplitudes can be calculated for high energies from the underlying Higgs-Goldstone system. 
The resulting $S$ matrix is diagonalized numerically and the imposed condition is for the largest
eigenvalue to be less than $8\pi$ in absolute value.
{ 
To help restrict further the parameter space we impose also unitarity conditions at finite energy $\sqrt{s}$, in which the full
tree level (scalar) two particle scattering matrix is constrained in an analogous fashion as in the LQT case but for an entire
range of scattering energies above the weak scale. This procedure not only strengthens the constraints over the quartic couplings
but also introduces new ones over the trilinear scalar couplings, which can be important
(as demonstrated recently in \cite{Goodsell:2018fex,Krauss:2018orw}).

It is convenient to have as many physical observables as possible serving as free parameters to increase the efficiency of the scanning 
algorithm, for this purpose it is straightforward to invert equations (\ref{alfa}) through (\ref{masses}) to work with the set of free parameters given by
all the physical masses plus the set $\mu_2^2$, $\lambda_{13}$, $\lambda_{14}$, $\tan{\theta}$ and $\alpha$, the only subtlety arises when manipulating
equations (\ref{alfa}) and  (\ref{mHh}) where there are two solutions for ($\lambda_1$, $\lambda_5$, $\lambda_8$) in terms of ($m^2_H$, $m^2_h$, $\alpha$)
and care must be taken to choose the one consistent with $(m^2_H-m^2_h) \cos{2\alpha} = {\cal M}_{11} - {\cal M}_{22}$ which follows from the same 
equations\footnote{We also take $\cos{2\alpha}>0$ to ensure the correctness of the diagonalization (\ref{zeta}) which is only valid for $|{\cal M}_{11} - {\cal M}_{22}| - ({\cal M}_{11} - {\cal M}_{22}) = 0$.}.

}
\subsection{The matter sector}

In the matter sector, invariance under the $S3$ symmetry implies a lagrangian of the form:

 \begin{equation}
-{\cal L}_Y = Y^{\prime Ds}_{\alpha\beta} \bar{Q}^\prime_{\alpha L} \Phi_s q^{\prime D}_{\beta R} + \\
              Y^{\prime D1}_{\alpha\beta} \bar{Q}^\prime_{\alpha L} \Phi_1 q^{\prime D}_{\beta R} + \\
              Y^{\prime D2}_{\alpha\beta} \bar{Q}^\prime_{\alpha L} \Phi_2 q^{\prime D}_{\beta R} + ...
 \end{equation}

where the transformation properties of the matter fields under $S3$ are taken as in reference \cite{Canales:2013cga}.

Here $D$ denotes $d$-type quarks and the dots refer to similar expressions for $u$-type quarks and leptons plus hermitian conjugate.
Flavor indices are denoted by $\alpha$ and $\beta$, quark left doublets and right singlets are denoted by
$Q_L$ and $q_R$ respectively, while unprimed quantities will refer to the mass eigenstate basis. The Yukawa matrices are given by:

\begin{eqnarray}
Y^{\prime D1} = 
\left(
\begin{array}{ccc}
 0 & \frac{y_{d3}}{2} & \frac{y_{d5}}{\sqrt{2}} \\
 \frac{y_{d3}}{2} & 0 & 0 \\
 \frac{y_{d6}}{\sqrt{2}} & 0 & 0 \\
\end{array}
\right)
&  ,  &
Y^{\prime D2} =
\left(
\begin{array}{ccc}
 \frac{y_{d3}}{2} & 0 & 0 \\
 0 & -\frac{y_{d3}}{2} & \frac{y_{d5}}{\sqrt{2}} \\
 0 & \frac{y_{d6}}{\sqrt{2}} & 0 \\
\end{array}
\right)
\end{eqnarray}
 and 

\begin{equation}
Y^{\prime Ds} = \text{diag}(\frac{y_{d2}}{\sqrt{2}}, \frac{y_{d2}}{\sqrt{2}}, y_{d1} )
\end{equation}
and similar expressions for $u$-quarks and leptons. 

It is well known from the studies of multi-Higgs models that the coupling 
of the fermions to extra scalar doublets can induce flavor changing currents even at tree level. Nevertheless, it has been shown in 
previous research concerning the $S3$ model (e.g. \cite{Canales:2013cga,Mondragon:2007af}) that values of the parameters $y_{d1}, \ldots$
calculated from fits to the CKM and PMNS mixing matrices are naturally small, so that experimental bounds on flavor changing
processes are not violated. For our purposes, we will assume in the following that the off-diagonal Yukawa couplings, given its
small size, do not contribute noticeably to the physical processes calculated in the next section, and thus we simply take this couplings
identically zero. Writing explicitly the quark fields for each family, with these assumptions after EWSB 
the parts in the Yukawa lagrangian involving neutral scalar fields become:

\begin{equation}
-{\cal L}_Y = y_{d1} (\bar{b}^\prime_{L} h^{\prime n}_s b^\prime_{R}) +
              \frac{y_{d2}}{\sqrt{2}} ( \bar{d}^\prime_{L} h^{\prime n}_s d^\prime_{R} +  \bar{s}^\prime_{L} h^{\prime n}_s s^\prime_{R}) + ... 
\end{equation}
After rotating the scalars to the mass eigenstate basis using (\ref{rotM}) we find

{\begin{equation}
h^{\prime n}_s =  H \, \cos{\alpha}  - h \, \sin{\alpha} 
\end{equation}
and thus $H_3$ does not couple to the fermions in this limit.
}
In the numerical analysis
we make the approximation of massless fermions for the first two families since the contributions
to the relevant observables are dominated by the masses of the third family of fermions.

\section{Numerical analysis and results}

\begin{figure}[t]
\includegraphics[width=0.5\linewidth]{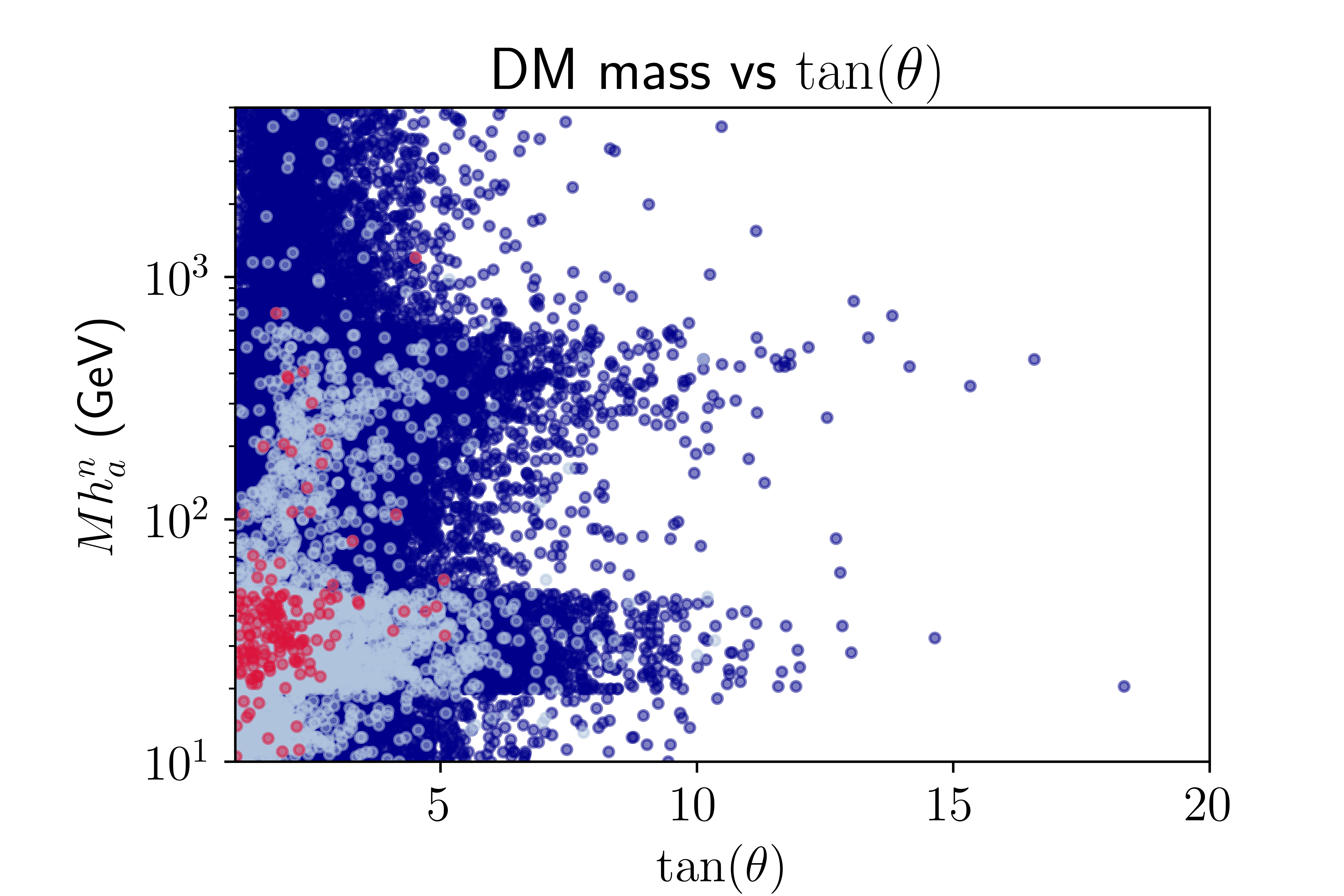} 
\includegraphics[width=0.5\linewidth]{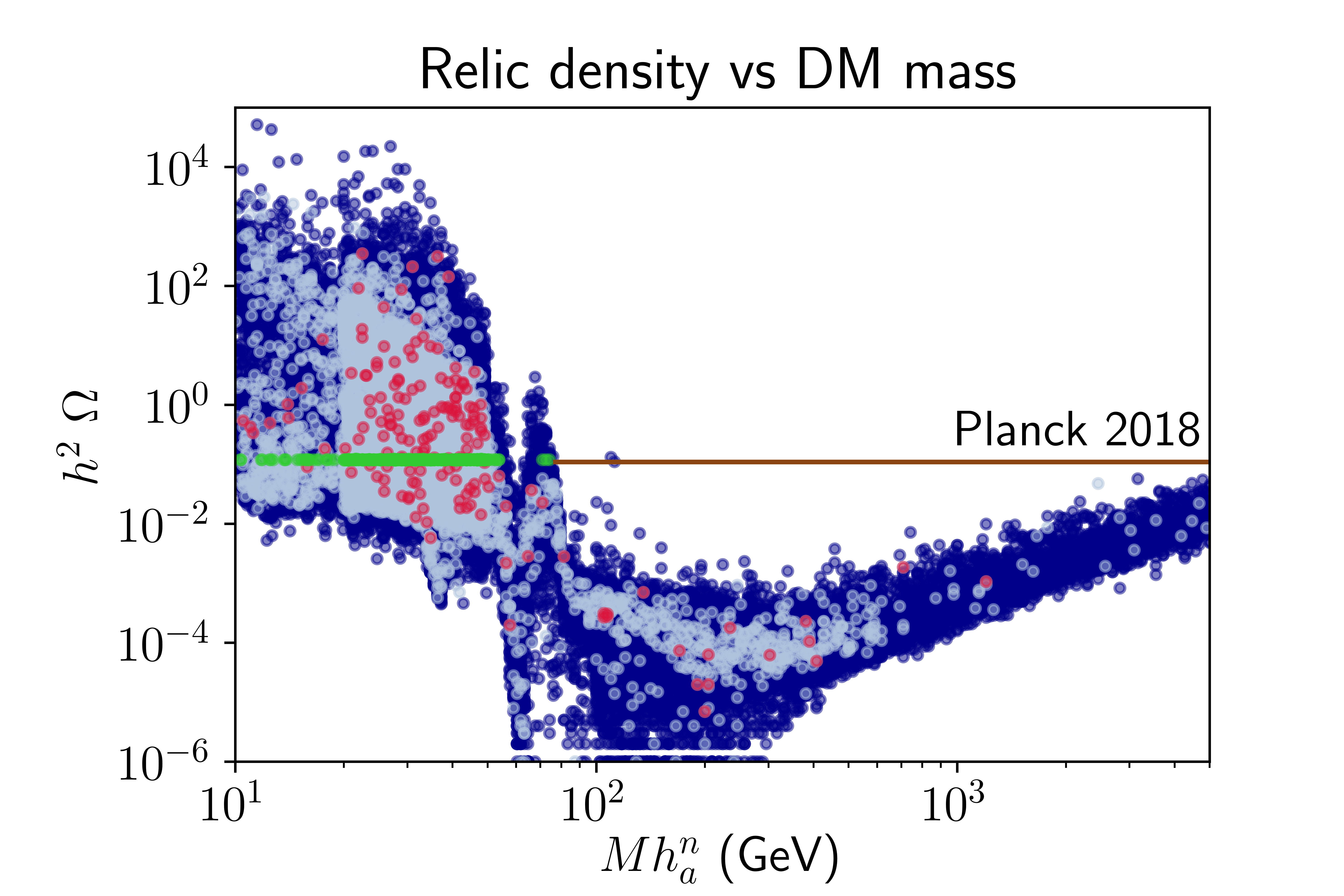} 
\caption{\label{DM_tan} Mass of the DM candidate as a function of $\tan\theta$ (left panel), and value of the DM relic density as a function of the DM mass. 
The dark blue points (set A) are the ones that comply with stability and unitarity constraints, the light blue points (set B) are also compatible with the experimental bounds for extra scalar searches (see text), the red points also satisfy the decoupling limit and the green points in the right panel lie within the experimental Planck bound.}
\end{figure}


\begin{figure}[h]
\includegraphics[width=0.5\linewidth]{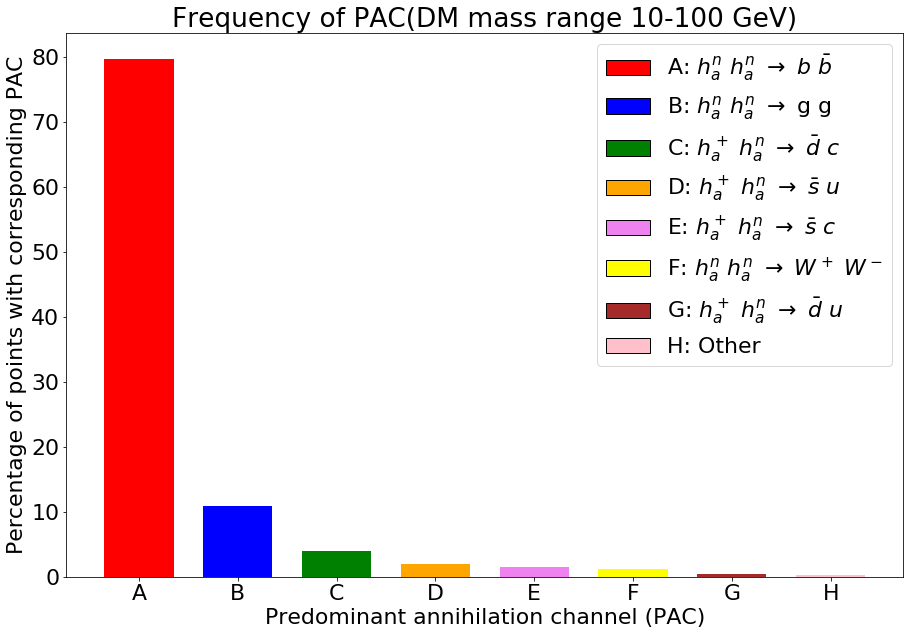}
\includegraphics[width=0.5\linewidth]{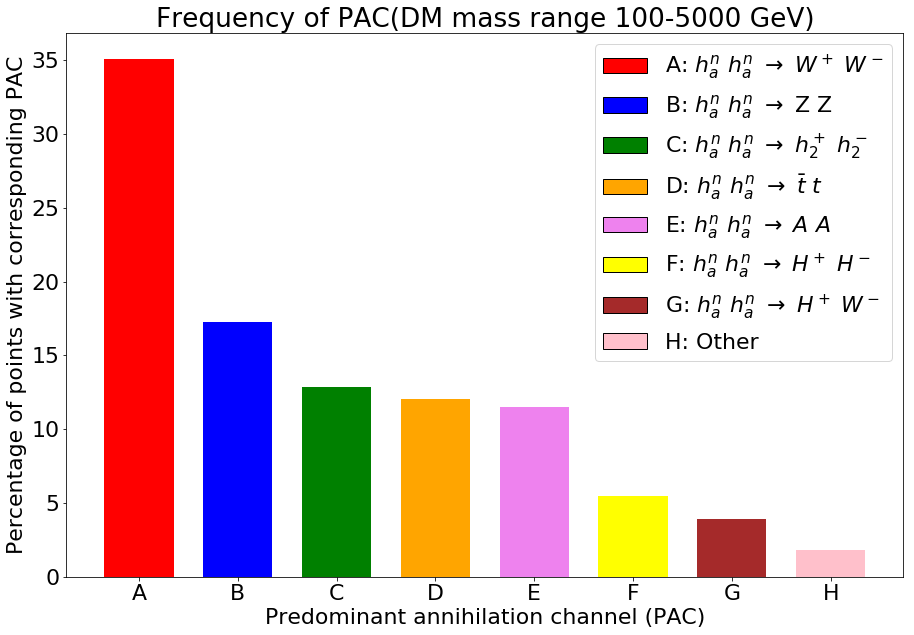}
\caption{\label{channels} Histograms of the frequency of predominant annihilation channels (for a full description see text).
}
\end{figure}

The implementation of the model is made through  \texttt{SARAH}~\cite{Staub:2013tta,Staub:2009bi,Staub:2010jh,Staub:2012pb} 
taking advantage of this package's functionality to generate model files 
for other tools. We perform a random scan of the parameter space filtering points that  do not satisfy all the conditions mentioned before;
the generation of the scattering matrix to calculate unitarity constraints in the large $s$ limit is done using \texttt{FeynArts}~\cite{Hahn:2000kx} 
and \texttt{FormCalc}~\cite{Hahn:1998yk}, 
{  whereas the calculation of the finite energy unitarity constraints is done with the latest \texttt{SARAH} update~\cite{Goodsell:2018tti}
only that we chose to port this specific part of the  \texttt{SARAH}-generated \texttt{SPheno} code to our scanning module for purposes of 
optimization\footnote{A slight optimization per point is gained by only executing \texttt{SPheno} for points that comply with all unitarity constraints, this
augments the efficiency of the code particularly when large amounts of points are being probed.}; for the handling of $t$ and $u$ poles in the calculation
of the scattering amplitudes, with hindsight we chose the weakest limits described in~\cite{Goodsell:2018tti} since already for this choice finding
physical points is computationally very expensive; the energy interval defined for these computations is taken to be $500$ to $5000$ GeV.
}
The generation of SLHA~\cite{Skands:2003cj,Allanach:2008qq} input files for 
\texttt{HiggsBounds}~\cite{arXiv:0811.4169,arXiv:1102.1898,arXiv:1301.2345,arXiv:1311.0055,arXiv:1507.06706}
and \texttt{MicrOMEGAS}~\cite{Barducci:2016pcb} is done using the 
\texttt{SARAH-\allowbreak SPheno}~\cite{Staub:2015kfa,Porod:2003um,Porod:2011nf} framework.
We use \texttt{HiggsBounds} to further filter points that do not comply with current experimental limits from Higgs searches, and 
finally \texttt{MicrOMEGAS} is utilized to compute the value of the relic density and annihilation cross section of the dark particle 
(the lightest of the $Z_2$-odd neutral scalars) for points that satisfy all the
constraints. 
{  We only show results for the case where the dark scalar $h_a^n$ is the dark matter candidate and we take its mass in the range $10$ to $5000$ GeV; similar conclusions are obtained
when the candidate is the pseudo-scalar $h_a^p$. All other dark particle masses are taken randomly in the range $\gtrsim M_{h_a^n}$ to $\sim 5000$ GeV,
while the heavy scalar masses take values in the range $\gtrsim M_{h}$ to $\sim 5000$ GeV.
For the parameter $\mu_2^2$ due to the first equation in (\ref{masses}) we generate random values for it in the interval $(\sim (-M^2_{h_a^+}), \sim M^2_{h_a^+})$, this should be 
a large interval to probe and in any case the value of $\lambda_{10}$ will be limited by the unitarity bounds and we don't expect large differences if this interval
is enlarged. Furthermore, this parameter also must satisfy the conditions (analogous to the IDM) $\mu_2^2 / \sqrt{\lambda_{13}} > \mathrm{Max}\{\mu_0^2 / \sqrt{\lambda_{8}},\mu_1^2 / \sqrt{\lambda_{1}}\}$ in order to prevent the possibility of tunnelling to a $v_a \neq 0$ vacuum~\cite{Ginzburg:2010wa}, to achieve this
we tune the value of $\lambda_{13}$ until the inequalities are satisfied, note that unphysical values of $\lambda_{13}$ will always be casted out by the subsequent filters.
Finally the values of the rest of the free parameters are taken in the ranges $\lambda_{14} \in [-4\pi, 4\pi]$, $\tan{\theta}\in (0, 100]$ and $\alpha \in [-\pi/4, \pi/4]$.
We present our results in figures (\ref{DM_tan}) through (\ref{xsect}).

The first observation that we want to make is that from the entire sample of points probed, those that satisfy the first line of constraints (close to $10^5$)\footnote{Note
that the total number of scanned points is far greater than this number since many of the randomly generated points are already discarded at the first line of constraints.}
i.e. stability and unitarity conditions,
only a small proportion (around $10\%$) passed the \texttt{HiggsBounds} tests. 
While the size of this sample\footnote{Larger sample sizes can 
easily become too expensive in computational time terms.}
 is rather small compared to 
the size of a parameter space of such dimensionality ($15$ total free parameters), we believe that the main conclusions drawn from our 
findings show important properties of the model. 

In the left panel of figure (\ref{DM_tan}) we present a scatter plot of the points in our scan projected in the plane DM mass vs $\tan{\theta}$. We refer to the points that pass the first line of constraints as set $A$ (dark blue points), while set $B$ (light blue points) are points that additionally satisfy the \texttt{HiggsBounds} limits; the subset of $B$ that satisfy the decoupling limit are shown in red color. From this figure we observe that stability and unitarity constraints severely restrict the values of $\tan{\theta}$ to $\lesssim 10$, with a handful
of points reaching up to 19, while the vast majority of the points in set $B$ lie below the 700 GeV mass mark. We also show in the right panel of this figure the value of the relic
density for each point as a function of the dark matter candidate's mass, with the color code for the points identical to the previous case except that points in set $B$ within the measured experimental value of Planck~\cite{Aghanim:2018eyx} are highlighted in green. 
From this figure we draw attention to the fact that for a large amount of the potential physical points found the dark scalar doublet of the model contributes only a fraction of the
experimentally measured value of the relic density, particularly in the region of masses above 80 GeV where we found no points in or above the Planck value.
On the other hand for masses below 80 GeV plentiful of the points shown have values of the relic well above the Planck limit and thus are unphysical; but
there are also points in and below this bound, the latter are not necessarily unphysical since there could be other sources of dark matter not taken into
account by the model.

In figure (\ref{channels}) we show histograms of the frequency of the dominant annihilation channels that contribute to the value of the 
relic density for each point of the scan in set $B$, where we distinguish the mass region where all points are below the Planck bound from the region
of small masses where points are found within the experimental interval. 
The color keys defined in the histograms display the percentage of points in the corresponding sets with given predominant annihilation channel (PAC),
defined as the channel with the highest branching ratio for each particular point of parameter space.
For the small mass range the dominant channel is by far annihilation into $b$ $\bar{b}$ pair since almost 80\% of the points
prefer this channel; interestingly we also see a few points where co-annihilation with $h_a^+$ is important.
For the high DM mass range (right panel histogram) around 50\% of the points annihilate into gauge bosons pairs and we find that
around a third of the points annihilate predominantly into pairs $h_2^+$ $h_2^-$, $t$ $\bar{t}$ and $A$ $A$ equally; and we don't find instances
where co-annihilation with other particles is important.

\begin{figure}[ht] \label{xsect}
\centering
\includegraphics[width=0.8\linewidth]{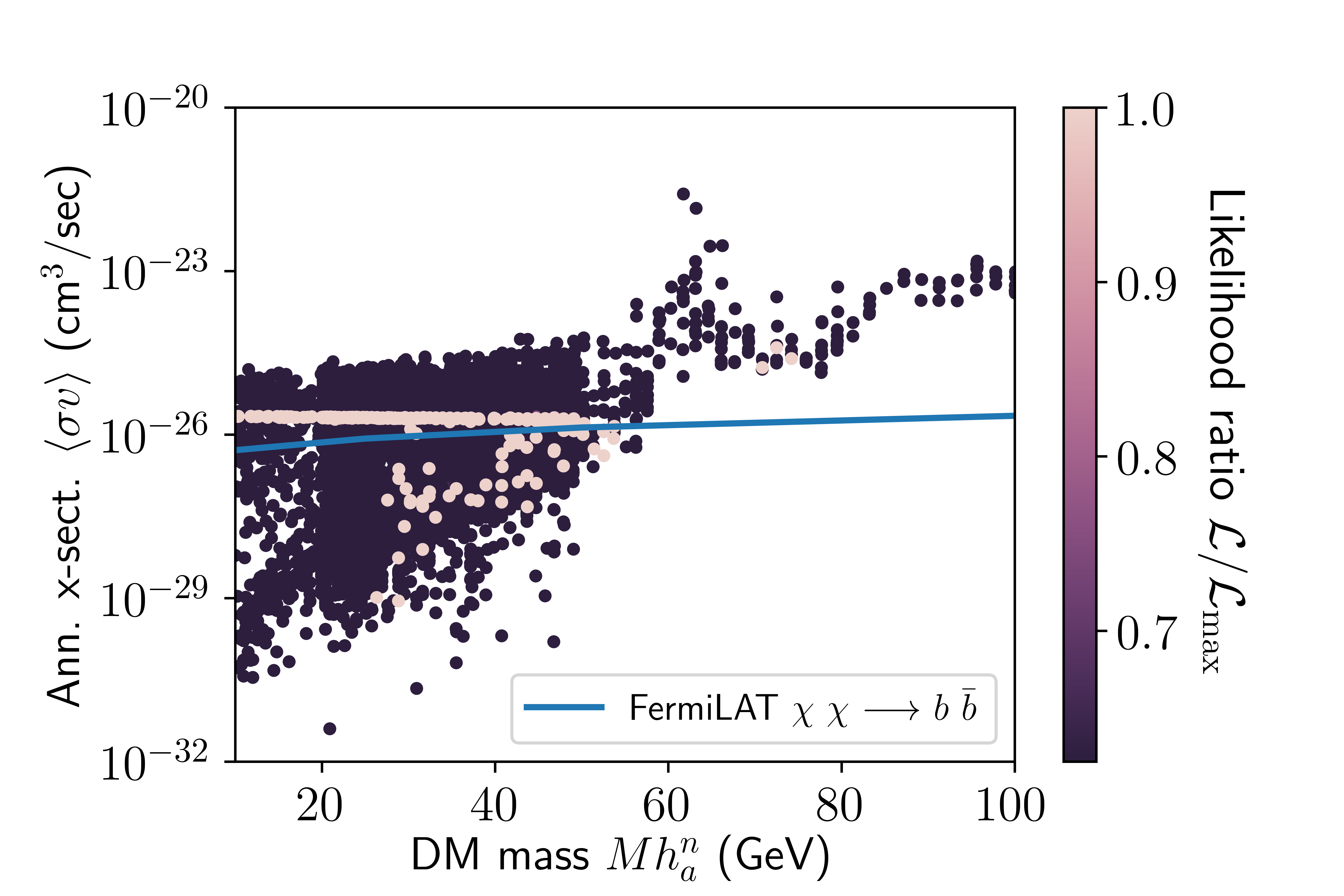}
\caption{\label{nmsn} Annihilation cross section as a function of the DM mass for small DM masses, the points are colored according to their
(normalized) likelihood (with respect to the relic density) value. Also shown is the FermiLAT dwarf spheroidal combined DM exclusion curve.}
\end{figure}

From the above results we deduce that the dark matter relic density dependence with the mass of our candidate follows a similar pattern to the i2HDM~\cite{Belyaev:2016lok}. Since we have more scalars than in the i2HDM there will be more decay channels available, but similar features remain. The left part (low mass region) will be dominated by Higgs mediated diagrams in the s-channel decaying into light fermions, then, there is a sharp dip where the candidate annihilates through the SM Higgs boson,
 for the most part this is a consequence of the resonance around the value of the Higgs mass due
to the s-pole in the annihilation amplitude
(for example two DM particles annihilating at rest will hit the resonance at around a mass of 62 GeV and thus the location of the dip in the figure); 
note that resonances due to the additional diagrams with heavy scalars do not appear as sharp dips because their masses are not fixed like the SM Higgs mass.
Higher order corrections will shift the locations of the poles but these corrections are not taken into account by \texttt{MicrOMEGAS}.
In the large mass region  the quartic interaction and the s-channel decays into gauge bosons are dominant, followed by the quartic channel dominated decay into charged scalars, s-channel dominated decay into top-anti-top pair, and quartic channel dominated decay into pseudoscalars. With rising DM mass the 
interplay between these decays, together with the values of the effective coupling $\lambda_X^\pm$ and the heavy scalar masses, will lead to a slow increase of the relic density with the dark matter candidate mass. In our case this increase is less steep than in the i2HDM, and it reaches values close to the Planck bound for masses  $~\sim 5$ TeV.

Finally in figure (\ref{xsect}) we present the annihilation cross section (relevant for Indirect Detection Experiments) as a function of the DM mass
for points with masses below 100 GeV; to highlight the points that lie within the Planck bound we define a likelihood function ${\cal L}$ with respect to the value of the
relic density, namely a gaussian centered in the Planck value with width equal to the 68\% experimental interval~\cite{Aghanim:2018eyx}, the points are colored
according to their ${\cal L} / {\cal L_{\mathrm{max}}}$ value, where ${\cal L_{\mathrm{max}}}$ corresponds to the maximum value attained within the set of points.
Thus, the darkest points in this figure lie above or below the Planck bound whilst the light pink points have relic density values within the experimental limits.
In the figure we also show the FermiLAT combined limits from dwarf spheroidal galaxies~\cite{Ackermann:2015zua} for the $b$ $\bar{b}$ channel (for this range of
masses other channel limits are indistinguishable). We see that the FermiLAT exclusion curve rules out many of the points compliant with the value of the experimental
relic density, but still some of these points lie safely below this bound. 

}

\section{Conclusions}

We have performed an analysis of the $S3$ symmetric model where the number of scalar doublets fill in all the irreducible
representations of the permutation group; by taking one of the scalars as inert we are retaining convenient features of the 
3-Higgs Doublet Model with $S3$ symmetry whilst additionally extending the model with a dark sector. We have probed random samples 
of parameter space points and find that the combination of physical constraints severely limits the parameter space,
in particular the most stringent constraints come from current experimental limits on Higgs searches. 
For a set of potentially physical points found in the scan we calculated the value of the relic density of the
dark scalar. 
The results obtained are similar to the i2HDM. There are points that comply with the Higgs searches bounds all along the dark matter mass spectrum probed. There is a low dark matter mass region $10 \sim 100$ GeV with points that have an overproduction of dark matter, but others that comply with the Planck bound and some that are below it. In this small mass region there are also points that satisfy the scalar searches bounds and relic density abundance Planck bound, as well as the FermiLAT combined limits for the annihilation cross section for the $b \bar{b}$ channel.
For the rest of the mass values there is an underproduction of dark matter due to an interplay between the annihilation channels,  the values of the effective parameter $\lambda ^\pm_X$, and the heavy scalar masses. In this case, for very heavy masses $\sim 5$ TeV there are some points that comply with the Higgs searches and which have a DM relic density close to the Planck measurement.

Our results suggest the possibility of extending the dark sector of the model with additional particles, potentially enriching
the phenomenology of this type of models. The inclusion of loop corrections to our results might change the outcome of the analysis, especially in the Higgs sector.

\section*{Acknowledgements}

We thank F. Staub for kindly clarifying in the \texttt{SARAH} forum many aspects of the \texttt{SARAH-SPheno} framework.
We also thank an anonymous referee for pointing out the existence of unitarity conditions at finite energy, as well as for other helpful suggestions.
C.E. acknowledges the support of CONACYT (M\'exico) C\'atedra 341. This work is supported in part by grant
PAPIIT project IN111518.


\section*{References}
\bibliography{darkS3}

\end{document}